\documentclass[aps, prl, 10pt, twocolumn, superscriptaddress, noshowpacs, preprintnumbers, floatfix]{revtex4-2}

\usepackage[colorlinks=true,breaklinks=true]{hyperref}
\hypersetup{allcolors=[rgb]{0.05 0.05 0.75},linkcolor=[rgb]{0.75 0.05 0.05}}
\usepackage[dvipsnames]{xcolor}
\usepackage{mathtools}
\usepackage{amsmath}
\usepackage{amsfonts}
\usepackage{amssymb}
\usepackage{bbold}
\usepackage{bm}
\usepackage{orcidlink}
\usepackage{esint}
\usepackage{microtype}

\long\def\exclude#1{}

\newcommand{\edit}[1]{{\color{black}{#1}}}

\newcommand{\bp}{{\bf p}}
\newcommand{\br}{{\bf r}}
\newcommand{\bv}{{\bf v}}
\newcommand{\tpsi}{{\tilde{\psi}}}
\newcommand{\GF}{G_{\rm F}}

\begin{document}

\title{Quasi-linear theory of fast flavor instabilities}

\author{Damiano F.\ G.\ Fiorillo \orcidlink{0000-0003-4927-9850}}
\affiliation{Istituto Nazionale di Fisica Nucleare (INFN), Sezione di Napoli,
Complesso Universitario di Monte Sant’Angelo, Via Cintia, 80126 Napoli, Italy}
\affiliation{Gran Sasso Science Institute (GSSI), Viale Francesco Crispi 7, 67100 L’Aquila, Italy}

\author{Georg G.\ Raffelt
\orcidlink{0000-0002-0199-9560}}
\affiliation{Max-Planck-Institut f\"ur Physik, Boltzmannstr.~8, 85748 Garching, Germany}

\begin{abstract}
Dense neutrino plasmas can develop instabilities that drive collisionless flavor exchange, equivalent to the emission of flavomons, the quanta of flavor waves. We treat these waves
as independent linear degrees of freedom and develop a quasi-linear theory (QLT), including backreaction on the neutrino distribution and nonresonant neutrino--flavomon interactions, while neglecting wave--wave processes. In a statistically homogeneous, axisymmetric model, the saturated neutrino and flavomon distributions agree closely with periodic-box solutions of the original quantum kinetic equation. Our results provide the first theoretical interpretation for the numerical observations. They also show that flavomon phases have little influence on the final state, supporting QLT
as a framework for capturing flavor evolution without resolving nonlinear small-scale dynamics.

\end{abstract}

\maketitle

{\bf\textit{Introduction}}---In the high-density environments of core-collapse supernovae and neutron-star mergers, it is neutrinos of all flavors that transport energy and lepton number \cite{Bethe:1985sox, Mirizzi:2015eza, Burrows:2020qrp, Boccioli:2024abp, Janka:2025tvf, Jerkstrand:2025, Raffelt:2025wty}. While their efficiency depends strongly on energy and flavor, the impact of flavor conversion remains poorly understood, primarily because collective flavor evolution \cite{Duan:2009cd, Duan:2010bg, Tamborra:2020cul, Richers:2022zug, Johns:2025mlm,Raffelt:2025wty}, caused by neutrino-neutrino refraction~\cite{Pantaleone:1992eq}, thwarts a first-principles implementation. While true conversion, driven by flavor-violating neutrino mass mixing, is suppressed by matter refraction \cite{Wolfenstein:1979ni}, collisionless flavor exchange among neutrinos may cause significant effects, although specific conclusions for now depend on parametric numerical implementations \cite{Ehring:2023lcd, Ehring:2023abs, Ehring:2024mjx, Xiong:2024pue, Wang:2025nii, Wang:2025vbx, Wang:2025ihh, Lund:2025jjo,Akaho:2026kff}.

Refractive flavor exchange can be efficient even in the limit of vanishing neutrino masses~\cite{Sawyer:2004ai, Sawyer:2008zs, Izaguirre:2016gsx}. Recent works focused entirely on such ``fast flavor conversion,'' although mass-driven (``slow'') collective evolution may have been prematurely dismissed because it need not be slow at all~\cite{Fiorillo:2024pns,Fiorillo:2025ank,Fiorillo:2025kko,Fiorillo:2025gkw} (see also Refs.~\cite{Shalgar:2020xns,Padilla-Gay:2025tko}). We still restrict our discussion to the fast case, which is governed by a single dimensional scale, the neutrino interaction energy, $\mu=\sqrt{2}\GF n_\nu$. It sets the scale for collective flavor effects to be much smaller than relevant hydrodynamical scales. It is this hierarchy that hinders a straightforward numerical implementation.

In the mean-field approach, neutrino flavor is encoded in density matrices $\varrho_\bp$ that have the usual occupation numbers as diagonal entries. In the massless (fast flavor) limit, one actually studies the density matrices for lepton number (neutrinos minus antineutrinos), following a well-known quantum kinetic equation (QKE) \cite{Dolgov:1980cq, Rudsky, Sigl:1993ctk, Fiorillo:2024fnl, Fiorillo:2024wej}
\begin{equation}\label{eq:QKE}
    (\partial_t+\bv\cdot\partial_\br)\varrho_\bp=-i\sqrt{2}\GF\!\int\frac{d^3\bp'}{(2\pi)^3}\left[\varrho_{\bp'},\varrho_\bp\right](1-\bv'\cdot \bv)
\end{equation}
with Fermi constant $\GF$ and neutrino velocity $\bv$. One can solve this QKE in axial symmetry for 
periodic-box
setups and extract fitting prescriptions for the final flavor distribution \cite{Bhattacharyya:2020jpj, Zaizen:2022cik, Xiong:2023vcm, Richers:2024zit, Liu:2025tnf, Goimil-Garcia:2025ozm}, which can then be implemented in large-scale simulations~\cite{Xiong:2024pue, Wang:2025nii, Wang:2025vbx, Wang:2025ihh, Akaho:2026kff}. One key assumption of this strategy is locality: each fluid element relaxes independently.

With the ultimate goal of avoiding the numerical small-scale problem and going beyond purely local solutions, we have recently developed a new perspective on collective flavor evolution \cite{Fiorillo:2024bzm, Fiorillo:2024uki, Fiorillo:2024dik, Fiorillo:2024pns, Fiorillo:2025ank, Fiorillo:2025zio, Fiorillo:2025kko, Fiorillo:2024qbl, Fiorillo:2025npi}, inspired by traditional methods of plasma physics. The key ingredient is the independent role of $\varrho_\bp(t,\br)$ disturbances in the form of flavor waves and their quanta, flavomons. In a two-flavor setup, we~write
\begin{equation}\label{eq:rho-expansion}
    \varrho_\bp=\frac{1}{2}\begin{pmatrix}
        n_\bp+D_\bp&\psi^*_\bp\\
        \psi_\bp & n_\bp-D_\bp
    \end{pmatrix},
\end{equation}
where $n_\bp=(f_\bp^{\nu_e}-f_\bp^{\bar\nu_e})+(f_\bp^{\nu_\mu}-f_\bp^{\bar\nu_\mu})$ is the total occupation number; it does not enter the evolution. The flavor content is measured by the DLN, or electron--muon lepton number difference, $D_\bp=(f_\bp^{\nu_e}-f_\bp^{\bar\nu_e})-(f_\bp^{\nu_\mu}-f_\bp^{\bar\nu_\mu})$, whereas flavor coherence by the complex field $\psi_\bp$. 

In linear theory, this field $\psi_\bp(t,\br)$ is small and its quantized disturbances are what we call flavomons, whereas those of $D_\bp(t,\br)$ would be neutrino--plasmons. A positive imaginary part of the $\psi$ dispersion relation signifies an instability, leading to fast flavor conversion. Our new paradigm, that works particularly well for weak instabilities (small growth rates), interprets the instability as the stimulated emission of flavomons by an out-of-equilibrium neutrino distribution. The growth rate corresponds to the emission rate of flavor Cherenkov radiation~\cite{Fiorillo:2024qbl, Fiorillo:2025npi}. 

The next step is quasi-linear theory (QLT), extending this picture
into the nonlinear regime by including backreaction on the DLN distribution, i.e., the response of $D_\bp$ to flavomon emission and absorption \cite{Fiorillo:2024qbl, Fiorillo:2025npi}.  For strong instabilities, this approximation may break down, so that one could systematically move forward using the weak-turbulence formalism that includes wave--wave interactions as we have mentioned earlier~\cite{Fiorillo:2025npi}.

We stress that the role of flavor waves is here entirely classical and indeed, our earlier treatment of QLT was formulated as such~\cite{Fiorillo:2024qbl}. Still, we often prefer the notion of flavomons because it brings out more clearly the underlying physical processes and makes a kinetic treatment more transparent. Notice that flavomons carry flavor so that an individual neutrino can only emit an individual quantum by flipping its flavor~\cite{Fiorillo:2025npi}, however not precluding a classical treatment.

Subsequent to our earlier papers on \edit{QLT and} flavomon kinetics \cite{Fiorillo:2024qbl, Fiorillo:2025npi}, Johns and Kost~\cite{Johns:2025yxa} have advanced a similar proposal of treating flavor waves as emergent degrees of freedom. Their weak-inhomogeneity approximation relies on a similar linear treatment of flavor-wave fluctuations. They develop an ambitious vision of the subject, based on an expansion around a generic space-time varying neutrino background, which for the moment, however, remains mostly formal.

\begin{figure*}
    \vskip-2pt
    \includegraphics[width=\textwidth]{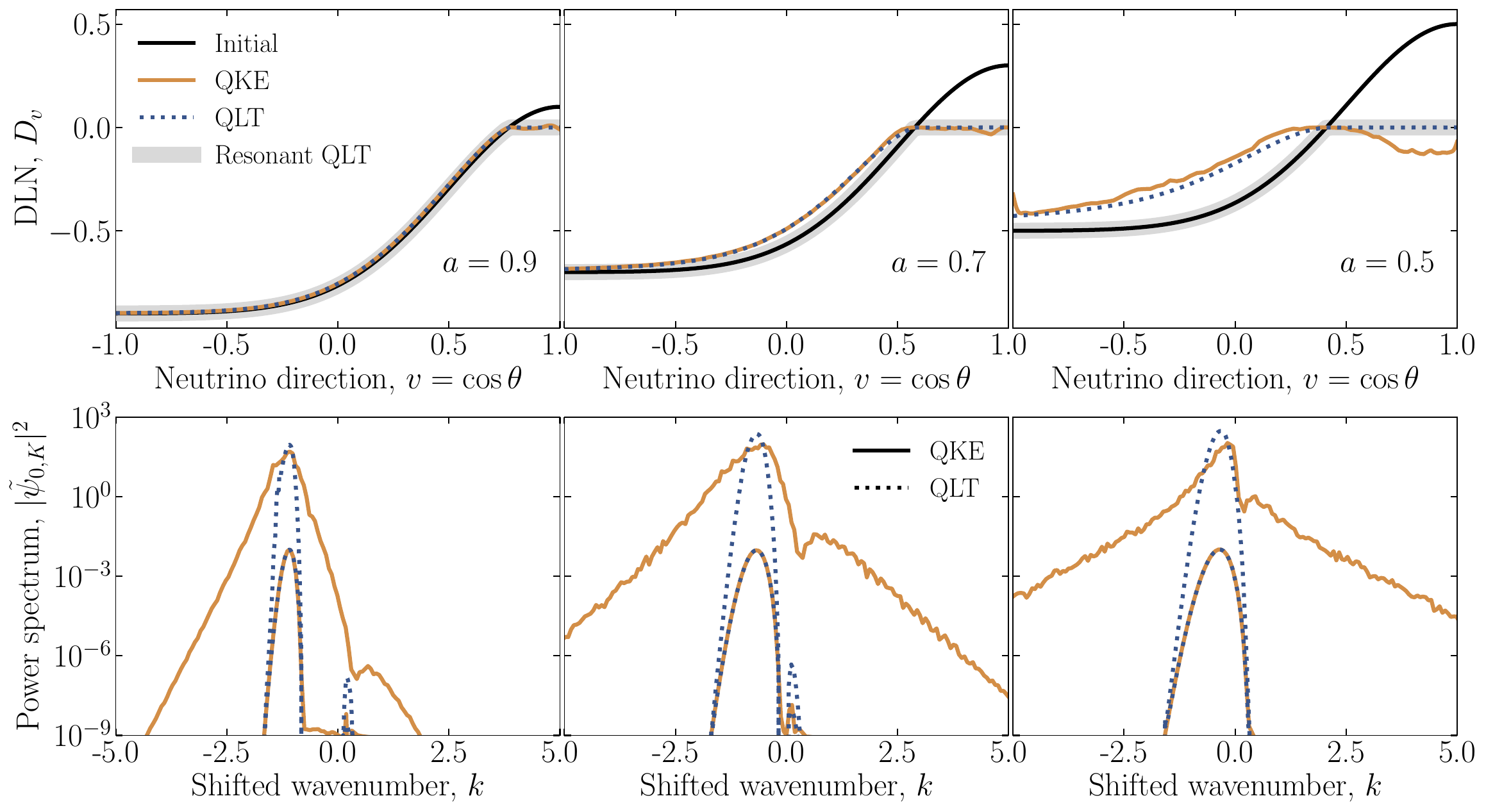}
    \vskip-6pt
    \caption{Comparison of quasi-linear theory (QLT) with simulations of the original quantum kinetic equation (QKE), averaged over the full periodic box, as well as over 20 realizations from the ensemble described in the text.
    \textit{Top:} DLN spectrum (electron--muon lepton number difference) derived with the indicated methods, with an initial distributions defined by Eq.~\eqref{eq:example_distribution}, for the shown values of $a$, left to right ranging from weak to strong instability. \textit{Bottom:} Final flavomon spatial power spectrum (upper lines) and at an intermediate time step (lower lines), chosen so that \smash{$|\tilde{\psi}_{0,K}|^2=10^{-2}$} at the maximum.
    }\label{fig:spectra}
    \vskip-8pt
\end{figure*}

In this Letter, we develop QLT beyond the limitations of our earlier approximations \cite{Fiorillo:2024qbl, Fiorillo:2025npi}. In particular, we include the nonvanishing flavomon width, allowing for non\-resonant absorption. In this way, neutrinos and flavomons reach self-consistent equilibrium that conserves global flavor lepton number. Moreover, we now treat the realistic case of a continuous distribution of unstable modes, tracking the power in each of them, but not the amplitude, going far beyond the earlier test case of a two-beam setup~\cite{Fiorillo:2024qbl}. As we will see, the saturated final state then matches surprisingly well
numerical periodic-box simulations of the original QKE.

\textit{\textbf{Axisymmetric homogeneous system}}---Following near-universal practice, we assume axial symmetry, so that $\varrho_\bp$ depends only on energy $|\bp|$ and $v=\cos\theta$ relative to the symmetry axis $z$. The fast flavor QKE does not depend on energy, motivating the integrated density matrix $\rho_v=n_\nu^{-1}\int \bp^2 d|\bp|/(4\pi^2)\,\varrho_\bp$, which obeys
\begin{equation}
    (\partial_t+v\partial_z) \rho_v=-i\mu\int_{-1}^{+1} dv'\,[\rho_{v'},\rho_v](1-vv').
\end{equation}
In component form, and after absorbing $\mu$ in the units of space and time, our final QKEs are
\begin{equation}\label{eq:standard_qke}
  \begin{aligned}
(\partial_t +v\partial_z) D_v&=\frac{i}{2}\left[(\psi_0-v\psi_1)\psi_v^*-(\psi_0^*-v\psi_1^*)\psi_v\right],
\\
(\partial_t +v\partial_z)\psi_v&=i\,[(D_0-vD_1)\psi_v-(\psi_0-v\psi_1)D_v],
\end{aligned}      
\end{equation}
with moments $D_n=\int dv v^n D_v$ and $\psi_n=\int dv v^n \psi_v$.

To determine the final quasi-steady state distribution, we use the initial DLN spectrum,
\begin{equation}\label{eq:example_distribution}
    D_v(t=0,a)=e^{-2(1-v)^2}-a.
\end{equation}
For $e^{-8}<a<1$, this distribution has an angular crossing and is unstable. In Fig.~\ref{fig:spectra}, we show three cases $a=0.9$, 0.7, and 0.5, ranging from weak to strong instability. For yet smaller $a$ values, the situation becomes specular and the flipped neutrinos, those with DLN opposite to the main beam, have negative~$v$.

\textit{\textbf{Resonant QLT prediction}}---In the simplest picture of flavomons~\cite{Fiorillo:2025npi}, they have a purely real frequency and can be emitted only by resonant neutrinos that fulfill the Cherenkov condition precisely. This zero-width approximation is particularly well justified for a weak instability, corresponding to a small population of flipped neutrinos. Only these can unflip (change their flavor) by flavomon emission. Without further ado, one thus expects that this process continues until there are no more flipped neutrinos, whereas the main beam remains unaffected. One therefore expects the final DLN distributions shown by thick gray lines in the top row of Fig.~\ref{fig:spectra}. The speed of this relaxation process would require an actual calculation, but the final outcome is unique.

The backreaction characteristic of QLT is reflected in the removal of the flipped portion of the distribution---the instability effectively eliminates its own source. On the other hand, global DLN is not conserved, as evidenced by the mismatch between the integrals under the black and thick gray curves. Extending beyond this approximation is the main goal of this work.

\textit{\textbf{Numerical periodic-box solution}}---As a benchmark for testing such predictions, we solve the QKE of Eq.~\eqref{eq:standard_qke} numerically in a periodic box of length $L=100$ (in units where $\mu=1$) with $N=700$ spatial grid points. The instability is seeded by a small initial value $\psi_v=\frac{1}{2}\psi_0$, taken independent of $v$, and expanded~as
\begin{equation}
    \psi_0(z,t=0)=\sum_{K} \psi_{0,K}(0) e^{i K z}.
\end{equation}
The discrete wavenumbers are $K_n=2\pi n/L$ with $n=-N/2,\ldots, +N/2$ in steps of~1. The Fourier amplitudes are initialized with a flat power spectrum $|\psi_{0,K}(0)|^2=10^{-9}/L^2$ with random phases. Concerning normalization, recall that the continuum Fourier component is $\tpsi_K= L\psi_K$ if $\psi_K$ is the corresponding Fourier series component on a box of length $L$.

We evolve the initial conditions for a time 500 and, for the final configuration, take an average of 20 realizations of our random initial conditions. While the time evolution exceeds the crossing time of the box, the quasi-stationary final configuration of our homogeneous setup should not be influenced much by the box~size. We stress that homogeneity, in the conventional sense implied by all forms of turbulence, here refers to \textit{statistical} homogeneity, which includes fluctuations that are distributed statistically in a homogeneous fashion.

We show the numerical solutions thus derived as orange lines in the upper panels of Fig.~\ref{fig:spectra} and the power spectra as solid lines in the lower panels. Especially for weak instabilities ($1-a\ll 1$), the numerical quasi-stationary DLN distributions (upper panels) show precisely the effect of eliminating the flipped part of the distribution that was explained by resonant QLT.

\textit{\textbf{Quasi-linear equations}}---We next turn to improved QLT predictions, anticipated in Fig.~\ref{fig:spectra} as dotted lines. To derive them, we expand $\rho_v(t,z)$ around a homogeneous background that is taken as the flavor-diagonal configuration, although more general formulations are possible~\cite{Johns:2025yxa}. We linearize in the flavor waves, $\psi_v(t,z)$, the off-diagonals in Eq.~\eqref{eq:rho-expansion}. The characteristic backreaction is included through slowly varying $D_v(t)$, which however are taken to be homogeneous. Neutrino-plasmons, disturbances of this field, are second order in $\psi$ and thus ignored.

We expand $\psi_v(t,z)$ into its spatial Fourier components
$\psi_{v,K}(t)$ and find from the second line of Eq.~\eqref{eq:standard_qke}
\begin{equation}\label{eq:fluct_evolution}
    \partial_t \psi_{v,K}+i(kv-D_0)\psi_{v,K}=-i(\psi_{0,K}-v\psi_{1,K})D_v,
\end{equation}
where $k=K+D_1$ is the shifted wavenumber. The formal solution for $\psi_{v,K}$ involves a memory integral over the past evolution of $\psi_{0,K}$. 

However, assuming that the background evolves on timescales much longer than the wave period, we can follow the adiabatic approximation in which each neutrino direction follows the corresponding eigenmode. If $\Omega$ is the eigenfrequency, this assumption yields
\begin{equation}\label{eq:eigenmode}
\psi_{v,K}
=\frac{\psi_{0,K}-v\psi_{1,K}}{\omega-kv}\,D_v,
\end{equation}
where $\omega=\Omega+D_0$. This expression is fully determined by $\psi_{0,K}$ if we use lepton-number conservation which implies $\psi_{1,K}=\Omega\psi_{0,K}/K$ \cite{Fiorillo:2024uki, Fiorillo:2020bvs}. We include here only the unstable modes, their amplitude determined by the initial condition described in the End Matter. Stable or Landau-damped modes remain negligible.

To include the backreaction on the homogeneous $D_v(t)$, we average the first line of Eq.~\eqref{eq:standard_qke}; the right-hand side becomes a sum over Fourier modes $\psi_{v,K}$. Passing to the discrete wavenumbers $K$ within the box, we find
\begin{equation}
    \partial_t D_v=\sum_{K}\mathrm{Im}\left[(\psi_{0,K}^*-v\psi_{1,K}^*) \psi_{v,K}\right].
\end{equation}
Substituting Eq.~\eqref{eq:eigenmode}, we obtain
\begin{equation}\label{eq:QL_ec_D}
    \partial_t D_v=-\Gamma_v D_v,
\end{equation}
where the damping rate is
\begin{equation}\label{eq:Gamma}
    \Gamma_v=\sum_{K}\frac{\gamma}{(\omega_{R}-k v)^2+\gamma^2}\left|1-\frac{(\Omega_{R}+i\gamma) v}{K}\right|^2|\psi_{0,K}|^2.
\end{equation}
Here $\Omega_{R}$ and $\omega_{R}$ are the real parts of the frequency and shifted frequency, $k=K+D_1$ is the shifted wavevector, and $\gamma$ is the growth rate for the unstable mode with wavenumber $K$. So all of $\Omega_R$, $\omega_R$, $\gamma$, and $k$ depend on~$K$. In contrast to the earlier periodic-box solution, no ensemble average is required because the statistical average over fluctuations is implied by QLT itself. \edit{We stress that this equation also directly coincides with the one obtained from the quantum theory of flavomons~\cite{Fiorillo:2025npi} (see e.g. Eq.~(21) therein), once the energy-conservation delta function is replaced by the Lorentzian broadened by the width $\gamma$.}
 
Physically, flavomon emission and absorption damps the DLN, since flavomon emission unflips a flipped neutrino, and vice versa. The coefficient $\Gamma_v$ coincides with the interaction rate derived in Ref.~\cite{Fiorillo:2025npi}, with the crucial difference that, instead of a delta function, we now include the Lorentzian which reflects the nonzero width of unstable modes and therefore includes off-resonant interactions. 

The adiabatic approximation itself may become suspicious when the unstable mode approaches stability and its phase velocity exceeds the speed of light: as shown in Ref.~\cite{Fiorillo:2024uki}, such superluminal modes generally develop a branch-root singularity into two branches of undamped modes. (A concern about singular points of the dispersion relation was also raised in Refs.~\cite{Liu:2025muc, Fiorillo:2026tee}.) However, our expression for $\Gamma_v$ reveals that, as soon as the mode becomes stable, its contribution vanishes, so that its singular behavior is irrelevant to the damping rate. In reality, the mode never becomes truly stable: upon instability saturation, the ambient
population of flavomons provides a nonzero wave--wave interaction rate~\cite{Fiorillo:2025npi, Johns:2025yxa}.

The evolution of the flavomon field is governed, within the same adiabatic approximation, by
\begin{equation}\label{eq:QL_ec_psi}
\partial_t |\psi_{0,K}|^2=2\gamma|\psi_{0,K}|^2.
\end{equation}
Equations~\eqref{eq:QL_ec_D} and \eqref{eq:QL_ec_psi} form a closed system describing the coupled evolution of neutrinos and a bath of linearly evolving flavomons, without having to solve the small-scale dynamics. The effect of QLT is entirely encoded in the DLN relaxation coefficient $\Gamma_v$.

Our QL method follows the flavomon number density, encoded in the classical wave action $|\psi_{0,K}|^2$, while ignoring rapid phase variations. In our simple setup, this is justified if there is only a single unstable eigenmode for each $K$. If there were multiple unstable modes with comparable growth rates, one could still disregard interference, provided their frequencies are well separated. The only potentially dangerous case occurs for very similar real eigenfrequencies and growth rates at the same~$K$; we are not aware of any such example.

The QL relaxation term actually respects global DLN conservation. As a proof, we integrate over directions
\begin{eqnarray}
    \kern-2em\partial_t D_0&=&-\sum_{K}|\psi_{0,K}|^2\,\mathrm{Im}\int\frac{dv D_v}{\omega-kv}\left|1-\frac{\Omega v}{K}\right|^2
    \nonumber\\
    &=&-\sum_{K}|\psi_{0,K}|^2\,\mathrm{Im}\left[I_0+\frac{|\Omega|^2I_2}{K^2}-\frac{2\Omega_R I_1}{K}\right],
\end{eqnarray}
where $I_n=\int dv\,D_v v^n/(\omega-kv)$. Modes satisfying the dispersion relation imply~\cite{Fiorillo:2024uki, Fiorillo:2020bvs}
$I_0=1+\Omega^2/(kK-\omega \Omega)$, 
$I_1=K \Omega/(k K-\omega\Omega)$, and
$I_2=-1+K^2/(kK-\omega \Omega)$.
Explicit replacement then shows that indeed $\partial_t D_0=0$.

So QLT preserves the total DLN in the neutrino sector. This is markedly different from resonant QLT, which neglects nonresonant flavomon absorption,  and only the total DLN of flavomons and neutrinos is conserved. In our new version of QLT, the DLN lost by resonant neutrinos is immediately redistributed among nonresonant ones. This difference comes from a common ambiguity, in that flavomons are collective excitations of the nonresonant neutrinos. The same ambiguity exists in plasma QLT for the bump-on-tail instability~\cite{vedenov1961stability, drummond1961non}, in which the total momentum of the electron plasma is conserved only if nonresonant plasmon absorption is included, otherwise the momentum transported by the plasmons should also be included.

{\bf\textit{Quasi-linear relaxation}}---To compare this theory with the earlier examples, we solve numerically Eqs.~\eqref{eq:QL_ec_D} and~\eqref{eq:QL_ec_psi} for the same initial spectrum that was used in the periodic-box solution of the original QKE. The initial fluctuation spectrum $|\psi_{0,K}|^2$ in QLT should be interpreted as the coefficient of the exponentially growing mode for every $K$. We extract it from the initial perturbation of the standard QKE (see End Matter).

Our periodic-box solution of the standard QKE involves a Gauss-Legendre discretized velocity grid with 100 points, and a spatial grid of 700 points. For the QL equations, we use the same velocity grid, and a discretized grid for the wavenumbers involving 600 points equally spaced between $k=-2$ and $k=2$. We use the shifted wavenumber, even though in principle the physical wavenumber should be used; this however does not practically impact the result---in weak instabilities, for which QLT is applicable, the global quantities such as $D_1$ barely change across the entire evolution. To actually solve the equations, we need to find the dispersion relation at every stage of the relaxation process, which requires considerable numerical effort. In a future more elaborate treatment, one would probably instead use our earlier asymptotic forms of the weak-instability dispersion relations~\cite{Fiorillo:2024uki, Fiorillo:2024pns, Fiorillo:2025zio}.

The results for the final DLN distribution are shown in the upper panels of Fig.~\ref{fig:spectra} as dotted lines. They agree remarkably well with the QKE solution, especially for weak instabilities, where QLT is best justified. Compared with the prediction of resonant QLT (thick gray lines), the final spectrum of the main beam is higher, reflecting the nonresonant absorption of flavomons that was not possible in resonant QLT. Global DLN conservation means that in each panel, the integrals under the black (initial) spectrum, under the orange line (final QLT), and under the blue dotted line (final QKE) are equal, as we have numerically confirmed. As stressed earlier, resonant QLT (gray lines) do not conserve DLN, but still capture the flat part, the removal of the flipped portion of the spectrum.

For the strongest shown instability, with $a=0.5$, the numerical QKE solution deviates more visibly from QLT. But even here, in principle outside the range where QLT applies, the qualitative behavior is well captured. Thus, this new method, without the need to account for nonlinear small-scale interactions, is extremely accurate in reproducing the standard QKE solution.

The flavor evolution not only affects the DLN distribution, but also leads to the production of flavomons, measured by the growth of the wave action
$|\psi_{0,K}|^2$, equivalent to flavomon number.
In the bottom panels of Fig.~\ref{fig:spectra}, we compare the predictions of the QKE and of QLT. Initially, during the phase of linear growth, they completely coincide,
as one would expect. The Fourier spectrum consists of two growing bumps, corresponding to two separate intervals of wavenumbers exhibiting an instability (see e.g.\ the general discussion in Refs.~\cite{Fiorillo:2024uki, Fiorillo:2024dik, Fiorillo:2025zio}). 

When the instability saturates, the QLT solution stops evolving, since flavomons and neutrinos have reached equilibrium while
wave--wave processes are excluded. Therefore, the final power spectrum always consists of the two bumps, with one much more prominent than the other. 

However, the QKE solution reveals that wave--wave processes do play a role, broadening the bump with an exponential decline on both sides. This intriguing finding may pave the way to a weak-turbulence treatment~\cite{zakharov2012kolmogorov} of the wave--wave interactions introduced in Ref.~\cite{Fiorillo:2025npi} (see also Ref.~\cite{Johns:2025yxa}), suggesting a direction for future work.

For now, we notice that the broadening of the power spectrum cannot be reproduced in QLT, yet the overall strength and wavenumber of the peak are both reproduced reasonably well. We also notice that the numerical power spectrum shows an intriguing dip that may coincide with the position of the second unstable bump. Whether this observation has any significance requires methods beyond QLT, so we leave this question for the future.

{\bf\textit{Discussion}}---We have applied QLT to the saturation of fast flavor instabilities for a continuous angular distribution, for the first time introducing the random phase approximation (in comparison to Ref.~\cite{Fiorillo:2024qbl}) and going beyond the resonant limit~\cite{Fiorillo:2025npi}, thus ensuring global DLN conservation. QLT thus becomes as predictive as conventional numerical approaches, while providing a transparent physical interpretation.

In a periodic-box setup, the computational effort is comparable to direct QKE solutions, as QLT requires solving the dispersion relation at each time step. However, this cost can be mitigated: during the linear phase, the eigenmodes are stationary, and accurate approximations for the growth rates are available~\cite{Fiorillo:2024bzm, Fiorillo:2024uki, Fiorillo:2024dik, Fiorillo:2024pns, Fiorillo:2025ank, Fiorillo:2025zio, Fiorillo:2025kko}. In our proof-of-principle implementation, we chose a brute-force approach, but significant optimization is possible.

Alternatively, one can avoid the dispersion relation by numerically evolving the linearized equations for the Fourier-mode amplitudes \cite{Liu:2026}; this also matches with the formal proposal in Ref.~\cite{Johns:2025yxa}. \edit{The main advantage of evolving the squared amplitudes, rather than the Fourier amplitudes themselves, is that we need not follow the phase, which changes over timescales of the order of $\mathrm{Re}(\Omega)^{-1}$. Instead, the squared amplitude only changes over timescales of the order of $\mathrm{Im}(\Omega)^{-1}$, which for weak instabilities is much longer \hbox{\cite{Fiorillo:2024pns,Fiorillo:2025ank,Fiorillo:2025zio,Fiorillo:2025gkw,Fiorillo:2025kko}} and thus more manageable. This is indeed the conventional approach in plasma physics. The main advantage} emerges beyond periodic boxes, when QLT naturally accommodates scale separation by removing the flavomon amplitude altogether. We thus anticipate that future applications will be to inhomogeneous and slowly evolving~systems.

Even for fast instabilities, evolving backgrounds can invalidate local approximations based on periodic-box solutions, as flavor waves retain memory of their past evolution~\cite{Fiorillo:2024qbl, Urquilla:2025idk} and through propagation may induce nonlocal effects~\cite{Fiorillo:2025ank, Johns:2025yxa, Fiorillo:2025gkw,Fiorillo:2026tee}. In principle, QLT can capture such effects, since the wave degrees of freedom are explicitly followed, and their propagation in inhomogeneous environments can be treated within a WKB framework. Whether QLT then maintains its quantitative accuracy remains to be seen.

The main advantage of QLT is conceptual. The resonant approximation explains the saturated state as a consequence of detailed balance: in regions dominated by flavor conversion, flavomon emission ceases once $D_v=0$, while outside the crossing region the distribution remains largely unaffected. Our full QLT treatment captures the nonresonant effects---flavomons emitted by flipped neutrinos are off-resonantly absorbed by unflipped ones. The saturated state predicted by QLT coincides with the erasure of angular crossing that was numerically found \cite{Nagakura:2022kic, Zaizen:2023ihz, George:2024zxz, Richers:2022bkd, Xiong:2023vcm, Richers:2024zit, Liu:2025tnf, Goimil-Garcia:2025ozm}. However, through our method, the outcome of collective flavor conversion becomes physically transparent rather than purely empirical.

This picture closely parallels the classical bump-on-tail instability in plasma physics~\cite{vedenov1961stability, drummond1961non}, where unstable wave emission leads to diffusion in momentum space and saturation through the formation of a velocity plateau. In our case, flavomons transport flavor rather than energy, and the evolution is governed by a DLN relaxation coefficient $\Gamma_v$ in place of a momentum diffusion coefficient. The vanishing of the DLN in the unstable region is therefore the direct analog of plateau formation in plasma QLT.

QLT relies on 
modes with many wavenumbers contributing incoherently to relaxation, causing irreversibility despite the absence of collisions.
This assumption can break down in highly symmetric configurations, such as flavor pendula \cite{Hannestad:2006nj, Johns:2019izj, Padilla-Gay:2021haz, Fiorillo:2023mze, Fiorillo:2023hlk, Liu:2025muc, Fiorillo:2026lyz, Fiorillo:2020vkc}, although such symmetries are unlikely to occur in astrophysical environments. A more relevant limitation arises for very weak instabilities, where the unstable spectrum is narrow and a single mode may temporarily dominate, leading to coherent evolution~\cite{Fiorillo:2020bvs}. Realistically,
this phase will be transient, eventually reverting to incoherent dynamics captured by~QLT.

Moreover, QLT provides direct access to the spectrum of unstable modes. While it reproduces the dominant wavenumber and overall power spectrum, deviations at late times (Fig.~\ref{fig:spectra}) arise from wave--wave interactions \cite{Fiorillo:2025npi}. Incorporating these effects represents a natural direction for future work. 

Another intriguing aspect that deserves future study
is the question of seeding. While one often
seeds inhomogeneous modes
at the same level as homogeneous ones, in reality there is much more power in large-scale than small-scale fluctuations. For fast instabilities, this issue is likely unimportant, since weak instabilities usually do not involve large-scale modes unless fine-tuned neutrino distributions are considered. Therefore, small-scale modes would always grow much larger than large-scale ones even if the latter start with a larger initial amplitude.

{\bf\textit{Acknowledgments}}---We thank Andrea Caputo for useful comments. DFGF acknowledges support by the TAsP (Theoretical Astroparticle Physics) project. GGR acknowledges partial support by the German Research Foundation (DFG) through the Collaborative Research Centre ``Neutrinos and Dark Matter in Astro- and Particle Physics (NDM),'' Grant SFB--1258--283604770, and under Germany’s Excellence Strategy through the Cluster of Excellence ORIGINS EXC--2094--390783311.

\bibliographystyle{bibi}
\bibliography{References}

\onecolumngrid
\newpage

\begin{center}
\textbf{\large End Matter: Results from linear theory}
\end{center}

\bigskip

\twocolumngrid

We briefly review some key results from the linear theory of fast flavor instabilities, primarily drawn from some of our earlier papers~\cite{Fiorillo:2024bzm, Fiorillo:2024uki, Fiorillo:2025npi}. We then use these expressions to fully deduce the quasi-linear relaxation in the resonant approximation, allowing us to express the final flavomon field in terms of the initial strength of the instability.

{\bf\textit{Initial conditions for QLT}}---As derived in the main text, the off-diagonal perturbations of the density matrix evolve according to 
\begin{equation}
    \partial_t \psi_{v,K}+i k v\psi_{v,K}-i D_0 \psi_{v,K}=-i(\psi_{0,K}-v\psi_{1,K})D_v,
\end{equation}
where $k=K+D_1$ is the shifted wave number. In the linear regime, $D_0$ and $D_1$ can be taken to be constant. 

To derive the evolution of a given initial condition, we may perform a one-sided Fourier transform
\begin{equation}
    \psi_{v,\Omega K}=\int_0^{\infty}e^{i\Omega t}\psi_{v,K}(t)\,dt.
\end{equation}
The inverse transformation is given by
\begin{equation}
    \psi_{v,K}(t)=\int_{-\infty+i \Lambda}^{+\infty+i\Lambda}e^{-i\Omega t}\psi_{v,\Omega K}\frac{d\Omega}{2\pi},
\end{equation}
where the integration is over a line parallel to the real axis, with $\Lambda$ large enough to pass above all singularities of the integrand function.

Applying the Fourier transform to the kinetic equation, we directly find a solution
\begin{equation}
    \psi_{v,\Omega K}=\frac{i \psi_{v,K}(t=0)}{\omega-kv}+D_v\frac{\psi_{0,\Omega K}-v\psi_{1,\Omega K}}{\omega-kv}.
\end{equation}
The evolution of $\psi_{1,\Omega K}$ is connected to that of $\psi_{0,\Omega K}$ through lepton number conservation in the form
\begin{equation}
    \partial_t \psi_{0,K}+i K \psi_{1,K}=0.
\end{equation}
This result directly descends from the continuity equation for the integrated lepton number; a dedicated discussion can be found in Refs.~\cite{Fiorillo:2024uki,Fiorillo:2020vkc,Fiorillo:2020bvs}.
Its temporal Fourier transform leads to
\begin{equation}
    \psi_{1,\Omega K}=-\frac{i\psi_{0,K}(t=0)}{K}+\frac{\Omega \psi_{0,\Omega K}}{K}.
\end{equation}
Therefore, we finally find
\begin{eqnarray}
    \kern-1em\psi_{v,\Omega K}&=&\frac{i\psi_{v,K}(0)}{\omega-kv}+\frac{\psi_{0,\Omega K} D_v}{\omega-kv}\left(1-\frac{v\Omega}{K}\right)
    \nonumber\\[1ex]
    \kern-1em&&\kern6em{}+\frac{i v D_v}{\omega -k v}\, \frac{\psi_{0,K}(0)}{K},
\end{eqnarray}
where for brevity we now shorten the specification $t=0$ to the argument in parenthesis. 

Integrating this equation over all directions, we finally find the solution
\begin{equation}
    \psi_{0,\Omega K}=\frac{i}{\varepsilon(\Omega)}\left[\int dv \frac{\psi_{v,K}(0)}{\omega-kv}+\frac{I_1 \psi_{0,K}(0)}{K}\right].
\end{equation}
In the denominator, we have introduced the dielectric function
\begin{equation}
    \varepsilon(\Omega)=1-I_0+\frac{I_1\Omega}{K}.
\end{equation}

When we perform the inverse Fourier transform to deduce the time evolution of $\psi_{0,K}$, the contour integration over $\Omega$ can be pushed down below the real axis, but only if $\varepsilon(\Omega)$ has no zero in the upper half-plane, i.e., no instability. Otherwise, if there is an unstable solution, the inverse Fourier transform collects a pole contribution that equals $-2\pi i$ times the residue of the integrand function. Since the pole is in the upper half-plane, this contribution is exponentially growing in time, and reads
\begin{equation}
    \psi_{0,K}^{\rm exp}(t)=\frac{e^{-i\Omega t}}{\partial\varepsilon/\partial \Omega}\left[\frac{I_1 \psi_{0,K}(0)}{K}+\int dv \frac{\psi_{v,K}(0)}{\omega-kv}\right],
\end{equation}
where the entire expression is to be evaluated for $\Omega$ equal to the frequency of the unstable mode. 

Hence, for any given initial condition, we discover that the coefficient of the exponentially growing term is not directly $\psi_{0,K}^{\rm QKE}(0)$, the initial condition for the QKE, but rather its projection over the unstable mode. For this reason, when we solve our QLT equations, rather than initializing them with a spectrum $\psi_{0,K}^{\rm QLT}(0)$ corresponding to the QKE case, we consider only the projection of the initial condition on the unstable mode, i.e.,
\begin{equation}
    \bigl|\psi_{0,K}^{\rm QLT}\bigr|^2
    =\bigl|\psi_{0,K}^{\rm QKE}\bigr|^2\,\frac{\left|\frac{I_1}{K}+\int dv\,\frac{\psi_{v,K}(0)/\psi_{0,K}}{\omega-kv}\right|^2}{|\partial\varepsilon/\partial\Omega|^2}.
\end{equation}
The dispersion relation itself, determining the unstable pole, reads $\varepsilon(\Omega)=0$. 

{\bf\textit{Resonant QLT}}---For weakly unstable configurations, corresponding to resonant instabilities, we may determine the growth rate directly from an expansion $\Omega=\Omega_R+i\gamma$ with $\gamma\ll \Omega_R$. For $\gamma=0$, the dielectric function is complex, due to the specific prescription of dealing with the singularity at $\omega=kv$ \cite{Fiorillo:2023mze, Fiorillo:2024bzm, Fiorillo:2024uki}. Therefore,
\begin{eqnarray}
    \kern-2em\varepsilon(\Omega_R+i\gamma)&=&\mathrm{Re}[\varepsilon(\Omega_R)]+\partial_{\Omega_R} i\mathrm{Re}[\varepsilon(\Omega_R)]\gamma
    \nonumber\\[1ex]
    &-&i\pi\int dv D_v\left(\frac{\Omega v}{K}-1\right)\delta(\omega-kv),
\end{eqnarray}
where we have neglected the derivative of the imaginary part with respect to $\Omega_R$, since the growth rate itself is rather small, and therefore so should be the imaginary part $\mathrm{Im}[\varepsilon(\Omega_R)]$. Hence, if we set to zero the imaginary part of this expression, we finally find the approximate expression for the growth rate
\begin{equation}
    \gamma=\frac{\pi}{\partial\mathrm{Re}\,\varepsilon/\partial\Omega_R}\int dv D_v \left(\frac{\Omega_R v}{K}-1\right)\delta(\omega_R - k v).
\end{equation}
This expression was derived in more detail in Ref.~\cite{Fiorillo:2024uki} by a similar approach, and in Ref.~\cite{Fiorillo:2025npi} by determining the rate of flavomon emission and absorption through the Feynman~rules.

This expression for the growth rate allows us to immediately understand the dynamics of quasi-linear relaxation in the resonant approximation. In this approximation, we associate with each neutrino velocity $v$ a resonant wavenumber and frequency satisfying $\omega_R=kv$. Since the growth rate in this approximation is assumed to be infinitesimal, we may neglect here the suffix to denote the real part. In this case the evolution of $|\psi_{0,K}|^2$ is given by
\begin{equation}
    \partial_t |\psi_{0,K}|^2=\frac{2\pi}{|k|\partial \varepsilon/\partial\Omega}\left(\frac{\Omega v}{K}-1\right)D_v|\psi_{0,K}|^2,
\end{equation}
where by the same approximation we also write $\mathrm{Re}\,\varepsilon\simeq \varepsilon$, and $v$ is interpreted as the resonant velocity. Meanwhile, the evolution of the DLN itself can also be obtained by the resonant approximation, taking the limit $\gamma\to 0$ in Eq.~\eqref{eq:Gamma}, so that
\begin{equation}\label{eq:resonant_diffusion}
    \partial_t D_v=-\frac{\pi}{\left|\partial_k\omega - v\right|}\left(1-\frac{\Omega v}{K}\right)^2D_v|\psi_{0,K}|^2.
\end{equation}
Assuming that the unstable eigenmode itself, measured by $\Omega$ and $K$, is not strongly affected, as is true for weak instabilities, these equations can be simply solved. They admit a conserved invariant, which is interpreted as the total lepton number which is conserved in the reaction $\overline{\nu}_e\to\overline{\nu}_\mu \psi$
\begin{eqnarray}
    \kern-3em&&2|\partial_k\omega-v|D_v+|k|\frac{\partial \varepsilon}{\partial \Omega}\left(\frac{\Omega v}{K}-1\right)|\psi_{0,K}|^2
    \nonumber\\[1ex]
    \kern-3em&&\kern8em{}=2|\partial_k \omega- v| D_{v}(t=0).
\end{eqnarray}
This result shows that the final squared amplitude $|\psi_{0,K}|^2$, which is reached when $D_v=0$, is directly determined by the initial resonant DLN value $D_v(t=0)$, i.e., the number of flavomons is simply the number of originally flipped neutrinos that have converted. Furthermore, after replacing $|\psi_{0,K}|^2$ in Eq.~\eqref{eq:resonant_diffusion}, we find
\begin{equation}
    \partial_t D_v=-\frac{\gamma(t=0)}{D_v(t=0)}D_v\left[D_v(t=0)-D_v\right].
\end{equation}
We thus find that in QLT, the timescale for saturation of the instability directly coincides with the initial growth rate. This result is valid, of course, for very weak instabilities, while for stronger instabilities, the wave--wave coupling acting after QL saturation will delay the formation of a true steady state.

\end{document}